\documentclass[showpacs,twocolumn,pre]{revtex4}
\usepackage{amsmath}
\usepackage{latexsym}
\usepackage{epsfig}
\usepackage{makeidx}
\usepackage{amsfonts}

\begin{document}

\title{Self-organized anomalous aggregation of particles performing nonlinear and non-Markovian random walk}
\author{Sergei Fedotov, Nickolay Korabel }
\affiliation{School of Mathematics, The University of Manchester, Manchester M13 9PL, UK }

\begin{abstract}
We present a nonlinear and non-Markovian random walk model for stochastic
movement and the spatial aggregation of living organisms that have the
ability to sense population density. We take into account social crowding
effects for which the dispersal rate is a decreasing function of the
population density and residence time. We perform stochastic simulations of
random walk and discover the phenomenon of \textit{self-organized anomaly}
(SOA) which leads to a collapse of stationary aggregation pattern. This
anomalous regime is self-organized and arises without the need for a heavy
tailed waiting time distribution from the inception. Conditions have been
found under which the nonlinear random walk evolves into anomalous state
when all particles aggregate inside a tiny domain (anomalous aggregation). We obtain power-law stationary
density-dependent survival function and define the critical condition for
SOA as the divergence of mean residence time. The role of the initial conditions in different 
SOA scenarios is discussed. We observe phenomenon of transient anomalous bi-modal aggregation.
\end{abstract}

\maketitle

\section{Introduction}

Aggregation of motile organisms is an example of ecological spatial
self-organization due to direct or indirect interactions between individuals
\cite{G}. Many organisms (cells, birds, mammals) have the ability to sense a
population density which leads to density-dependent dispersal \cite{O,Erik}.
This dependence can be explained by various mechanisms including (i)
competition that forces an individual to emigrate (positive
density-dependence), (ii) avoidance of crowded areas by individuals
(positive density-dependence), (iii) social crowding effects when certain
areas are attractive to many conspecifics (negative density-dependence) \cite%
{Erik}. Density-dependent dispersal can be regarded as a behavioral response
in which an individual changes its rate of jump due to sensing the mean
population density \cite{Men,Men3}. To model the density-dependent dispersal
and aggregation, various nonlinear\textit{\ }local and non-local
advection-diffusion equations have been used in the literature \cite%
{Turchin,Pe, Mai,Men3,Cates,Kopel}. Some living organisms like amoeboid
microorganism \textit{Dictyostelium discoideum }can interact indirectly.
They secrete a diffusible attractant (signalling molecules) to which
individuals respond chemotactically. They move towards regions of high
concentration of attractant and aggregate into a mound. On the population
level, the standard model for chemical interaction of species is a pair of
the advection-diffusion-reaction equations for the average population
density and concentration of signalling molecules. There is a huge body of
literature on chemotactic aggregation modeling (see, for example, \cite%
{sad2,EO,HP,Erban,Brenner,The}).

The microscopic theory of actively moving organisms is based on various
random walk models \cite{Men,Men3,Turchin,Pe,
Mai,Cates,Kopel,sad2,EO,HP,Erban}. One of the main characteristics of the
continuous time random walk (position jump walk) is the escape rate $\mathbb{%
T}$ from the point $x.$ For density-dependent dispersal models involving
direct interaction between individuals this rate depends on the population
density \cite{Men3}. Indirect interaction can be modeled by the dependence
of $\mathbb{T}$ on concentration of signalling molecules produced by
individuals \cite{sad2,EO,HP,Erban}. An important feature of such random
walk models is that they are Markovian. However, many transport processes
are non-Markovian for which the transport operators are non-local in time.
Examples include the L\'{e}vy walk that may accelerate aggregation \cite%
{Kopel2}, slow subdiffusive transport that may lead to the phenomenon of
anomalous aggregation \cite{Fed3}. The challenge is to take into account
both nonlinear density-dependent dispersal and non-Markovian anomalous
behavior \cite{Klafter,Kla,Men2}. Although some research has been done to
address the interplay between nonlinearity and non-Markovian effects \cite%
{Fed2}, it is still an open problem.

In this paper we consider a nonlinear and non-Markovian random walk and
propose an alternative mechanism of aggregation which we call the regime of
\textit{self-organized anomaly }(SOA). It leads to the collapse of a
standard stationary aggregation pattern and development of non-stationary
anomalous aggregation. By using Monte Carlo simulations we show that under
certain conditions particles performing a non-Markovian random walk with
crowding effects aggregate inside a tiny domain (anomalous aggregation). 
Important fact is that this anomalous regime is self-organized and arises 
spontaneously without the need for a heavy tailed waiting time distribution 
with an infinite mean time from the inception. 

\section{Nonlinear and non-Markovian random walk}

In this section we formulate our nonlinear and non-Markovian continuous time
random walk model. Instead of the waiting time probability density function
(PDF) we use the escape rate $\mathbb{T}(\tau ,\rho)$ that depends on the
residence time $\tau $ and the density of particles $\rho.$ Due to the
dependence of the rate $\mathbb{T}$ on the residence time $\tau,$ our model
is non-Markovian and should involve memory effects. Our intention is to take
into account nonlinear social crowding effects and non-Markovian negative
aging. We assume that the random walker has the ability to sense the
population density $\rho $. We model the escape rate $\mathbb{T}$ as a
decreasing function of the density $\rho (x,t)$ \cite{Erik,Men3}
\begin{equation}
\mathbb{T}(\tau ,\rho )=\frac{\mu\left( \tau \right) }{1+A\rho(x,t)},
\label{T}
\end{equation}%
where $A$ is a positive parameter. This nonlinear function describes the
phenomenon of conspecific attraction: the rate at which individuals emigrate
from the point $x$ is reduced due to the presence of many conspecifics. This
negative density-dependence can be explained by the various benefits of
social aggregation like mating, anti-predator aggregation, etc. \cite{Erik}.
The rate parameter $\mu \left( \tau \right) $ is a decreasing function of
the residence time (negative aging):
\begin{equation}
\mu \left( \tau \right) =\frac{\mu _{0}}{\tau _{0}+\tau },  \label{d}
\end{equation}%
where $\mu _{0}$ and $\tau _{0}$ are positive parameters. This particular
choice of the rate parameter $\mu \left( \tau \right) $ has been motivated
by non-Markovian crowding: the longer the living organisms stay in a
particular site, the smaller becomes the escape probability to another site.
Since the escape rate$\ \mathbb{T}(\tau ,\rho )$ depends on both residence
time $\tau $ and $t$ (indirectly through $\rho $) we can not define the
waiting (residence) time PDF. It can be done only for the linear case when $%
A=0.$ In this case the waiting time PDF $\psi \left( \tau \right) $ can be
defined in the standard way: $\psi \left( \tau \right) =\mathbb{T}(\tau
,0)\exp \left[ -\int_{0}^{\tau }\mathbb{T}(\tau ,0)d\tau \right] $ \cite{Cox}%
. The particular choice (\ref{d}) generates the power law distribution:
\begin{equation}
\psi (\tau )=\frac{\mu _{0}\tau _{0}^{\mu _{0}}}{(\tau _{0}+\tau )^{1+\mu
_{0}}}.  \label{wtden}
\end{equation}%
For the exponent $\mu _{0}<1,$ this waiting time probability density
function has infinite first moment which corresponds to anomalous
subdiffusion \cite{Klafter,Kla,Men2}. In this paper we choose $\mu _{0}$ as
\begin{equation}
\mu _{0}>1
\end{equation}%
for which the mean waiting time is finite for the linear case ($A=0).$ We do
not introduce the anomalous effects from the inception as its is done for a
classical theory of subdiffusive transport \cite{Klafter,Kla,Men2}.

Regarding the space dynamics, we consider the random walk in the stationary
external field $S(x)$ on the one-dimensional lattice with the step size $a.$
We should note that the extension for the two and three dimensions is pretty
straightforward. When the walker escapes from the point $x$ with the rate $%
\mathbb{T}(\tau ,\rho )$, it jumps to $x+a$ with the probability $p_{+}(x),$
and it jumps to $x-a$, with the probability $p_{-}(x).$ For the standard
chemotaxis models \cite{Angela}, it is assumed that the jumping
probabilities are determined by the chemoattractant concentration $S(x)$ on
both sides of the point $x$ as%
\begin{equation}
p_{\pm }(x)=\frac{e^{\beta S(x\pm a)}}{e^{\beta S(x-a)}+e^{\beta S(x+a)}}.
\label{ppppp}
\end{equation}%
In this way we introduce the bias of the random walk in the direction of the
increase of the external field $S(x).$ The positive parameter $\beta >0$ is
the measure of the strength of the bias. For small step size $a,$ one
can obtain the expressions for $p_{\pm }(x)$ in the continues case
\begin{equation}
p_{\pm }(x)=\frac{1}{2}\pm \frac{\beta }{2}\frac{\partial S}{\partial x}%
a+o(a).  \label{pppp}
\end{equation}

Because of the dependence of $\mathbb{T}$ on the residence time $\tau ,$ it
is convenient to define the structured density of particles $\xi (x,\tau ,t)$
at time $t$ such that $\xi (x,\tau ,t)\Delta x\Delta \tau $\ gives the
number of particles in the space interval $\left( x,x+\Delta x\right) $\
whose residence time lies in $\left( \tau ,\tau +\Delta \tau \right) $ \cite%
{Fed2,Cox,Vlad}. We consider initial conditions $\xi (x,\tau ,0)=\rho_{0}(x)\delta(\tau)$, 
for which all particles have zero residence time at $t=0$. The total density $\rho (x,t)$ is defined in the standard
way%
\begin{equation}
\rho (x,t)=\int_{0}^{t}\xi (x,\tau ,t)d\tau.  \label{den5}
\end{equation}%
The balance equation for the density $\xi (x,\tau ,t)$\ for $\tau >0$\ takes
the Markovian form
\begin{equation*}
\xi (x,\tau +\Delta \tau ,t+\Delta t)=\xi (x,\tau ,t)\left( 1-\mathbb{T}%
(\tau ,\rho )\Delta \tau \right) +o\left( \Delta t\right) ,
\end{equation*}%
where $1-\mathbb{T}(\tau ,\rho )\Delta \tau $\ is the survival probability
during $\Delta \tau $\ at point $x$. Since $d\tau /dt=1,$\ in the limit $%
\Delta t\rightarrow 0$\ we obtain the following equation
\begin{equation}
\frac{\partial \xi }{\partial t}+\frac{\partial \xi }{\partial \tau }=-%
\mathbb{T}(\tau ,\rho )\xi .  \label{dif1}
\end{equation}%
In what follows we use this equation to determine the conditions for the
self-orginized anomalous regime. The master equation for $\rho$ can be written as \cite{Men} 
\begin{eqnarray}
\frac{\partial \rho }{\partial t} &=&-i(x,t)+p_{+}(x-a) i(x-a,t)  \notag \\
&&+p_{-}(x+a) i(x+a,t),
\label{masterGEN}
\end{eqnarray}%
where $i(x,t)=\int_{0}^{t} \mathbb{T}(\tau ,\rho )\xi(x,\tau,t) d\tau$. 
In general, the expression for the total escape rate $i(x,t)$ is not known. 
In the Markovian case, when the rate $\mathbb{T}$ is independent of $\tau$, 
using Eq.\ (\ref{den5}) we obtain 
\begin{equation}
i(x,t)=\mathbb{T}(\rho)\rho(x,t).
\label{nonmarkovian_rate} 
\end{equation}%
For $\mu_0<1$ and only for linear case ($A=0$), the total escape rate takes the form \cite{Fed3}
\begin{equation}
i(x,t)=\left( \Gamma (1-\mu_0) \tau_{0}^{\mu_0}\right)^{-1} D_{t}^{1-\mu_0} \rho(x,t),
\label{nonmarkovian_rate2}
\end{equation}
where $D_{t}^{1-\mu_0}$ is the Riemann-Liouville fractional derivative.

\section{Stochastic simulations and self-organized anomaly}

\begin{figure}[t]
\centerline{
\psfig{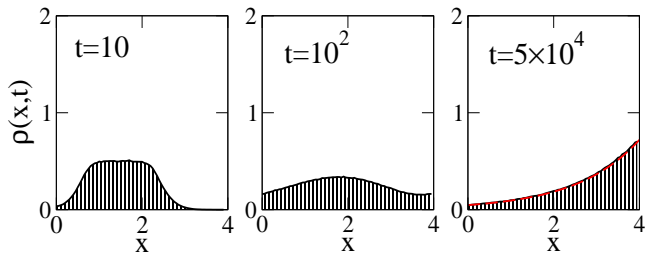}
}
\caption{Population density profiles in linear regime $A=0$. External field
(chemoattractant concentration) is linear: $S(x)=\protect\alpha x$.
Parameters are $\protect\tau_0=1$, $\protect\beta=1$, $\protect\alpha =0.34$
and $\protect\mu _{0}=4$. We consider $L=4$, and $M=40$, so that $a=0.1$.
The dashed line in the right most panel (coincides with the profile)
represents the stationary Boltzmann distribution. We have used ensemble of $%
10^{6}$ walkers uniformly distributed at time $t=0$ on the interval $%
[0.5,2.5]$ with zero residence time ($\protect\tau =0$). The boundaries at $%
x=0$ and $x=4$ are assumed to be reflecting.}
\label{FIG1}
\end{figure}

In this section we present the stochastic simulations of a nonlinear and
non-Markovian continuous time random walk along one-dimensional lattice.
Because of the density-dependent dispersal we intend to model, we can not
introduce the power law probability density function for the random
residence time as it is done for the classical continuous time random walk
(CTRW) \cite{Klafter,Kla,Men2}. Therefore we cannot apply the standard
Gillespie algorithm \cite{Gi}. In this paper, we simulate the random walk in
a domain $\left[ 0,L\right] $ which is divided into $M$ boxes of length $%
a=L/M$. To calculate the escape rate (\ref{T}) from the box, we approximate
the density of walkers $\rho $ in each box as the number of walkers in this
box divided by the size the box. Each walker has its own residence time $%
\tau $ which determines its escape rate (\ref{T}). First, we consider a
stationary linear profile
\begin{equation}
S(x)=\alpha x,  \label{linear}
\end{equation}%
where $\alpha $ is the chemotactic strength parameter. In Fig.\ \ref{FIG1}
we present a convergence of the population density profile $\rho (x,t)$ to
the stationary distribution $\rho _{st}(x)$ on the interval $[0,4]$ in the
linear case ($A=0)$ for which the walkers only sense the external field
(concentration of signalling molecules) $S(x)$ and not the density of
particles $\rho (x,t).$ One can see that the standard aggregation pattern develops.
It is easy to show that in the continuous case this distribution is the
stationary solution of the standard Fokker-Planck equation: $\rho
_{st}(x)\sim \exp \left[ 2\beta S(x)\right] .$

The striking feature of our random walk is that the interplay of
nonlinearity (\ref{T}) and non-Markovian aging effect (\ref{d}) leads to
non-equilibrium phase transition. In the nonlinear case ($A>0)$, when the 
chemotactic strength parameter $\alpha $ is smaller than a critical value $\alpha _{cr}$,
the population density convergences to the stationary aggregation profile. 
When the chemotactic strength parameter $\alpha $ is greater than $\alpha _{cr}$, 
crowding effects induce a collapse of the stationary
aggregation pattern. The nonlinear random walk evolves into the \textit{%
self-organized anomaly}. In this regime the stationary density profile does not exits and 
all walkers tend to aggregate at the point $x_{m}$ where the sub-critical ($\alpha<\alpha_{cr}$) stationary density $\rho _{st}(x)$ takes the
maximum value $\rho _{st}(x_m)$. The initial conditions are chosen such that $\rho_{0}(x) < \rho _{st}(x_m)$. 
We discuss the role of other initial conditions in section V. Stochastic simulations
of the nonlinear and non-Markovian random walk are presented in Fig. \ref%
{FIG2}. The top row shows the convergence of the population density to the
stationary aggregation profile for $\alpha =0.34$. When $\alpha <$ $\alpha
_{cr}\simeq 0.345,$ we observe a development of a stationary aggregation
with a continues increase in the maximum value of population density at the
point $x_{m}=4$ as the parameter $\alpha \ $\ increases up to $\alpha _{cr}.$
The value of $\alpha _{cr}$ depends on the properties of the random walk, 
system size and the boundary conditions. Here we do not study this dependence. 
The observed profile is similar to that of Fig.\ \ref{FIG1}. The only difference is
that the nonlinear crowding effects make the value of $\rho _{st}(4)$
greater. However, the drastic change happens, when the value of $\alpha $
exceeds $\alpha _{cr}.$ The dashed line in Fig.\ \ref{FIG2} represents $\rho
_{st}(4)=1.5$ given by Eq.\ (\ref%
{rho_cr}) below. Our nonlinear random walk evolves into the non-stationary
SOA which becomes an attractor for random dynamics. In this regime all particles eventually concentrate in the
vicinity of $x_{m}=4.$ The bottom row in Fig.\ \ref{FIG2} shows this collapse of the
density for $\alpha=1$. Note that similar phenomenon occurs as a result of
chemotaxis, the so-called \textit{chemotactic collapse} when all cells
aggregate at some point. Our explanation of this collapse is different from
the classical Patlak-Keller-Segel theory in which the growth of cell density
to infinity happens in finite time \cite{sad2,Brenner,The}.

%
\begin{figure}[t]
\centerline{
\psfig{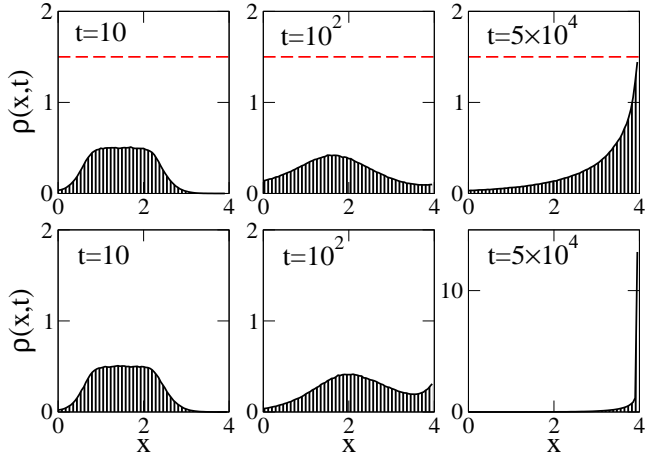}
}
\caption{Transition to self-organized anomalous regime for the linear
density of external field $S(x)=\protect\alpha x$. Transition from
stationary density to collapsed density occurs when signalling strength $%
\protect\alpha $ exceeds critical value $\protect\alpha _{cr}$, or in other
words when the maximum of the stationary density, $\protect\rho _{st}(x_{m})$%
, exceeds the critical value $\protect\rho _{cr}=1.5$ given by Eq.\ (%
\protect\ref{rho_cr}). Here $x_m=4$ and $\protect\mu _{0}=4$, $%
A=2$. Other parameters, initial and boundary
conditions are the same as in Fig.\ \ref{FIG1}. 
The top row shows the formation of stationary distribution for $%
\protect\alpha=0.34<\protect\alpha_{cr}$. The bottom row illustrates the
density collapse that takes place for $\protect\alpha=1>\protect\alpha_{cr}$%
. The critical strength of signalling is estimated numerically as $\protect%
\alpha _{cr}\simeq 0.345$.}
\label{FIG2}
\end{figure}

\begin{figure}[t]
\centerline{
\psfig{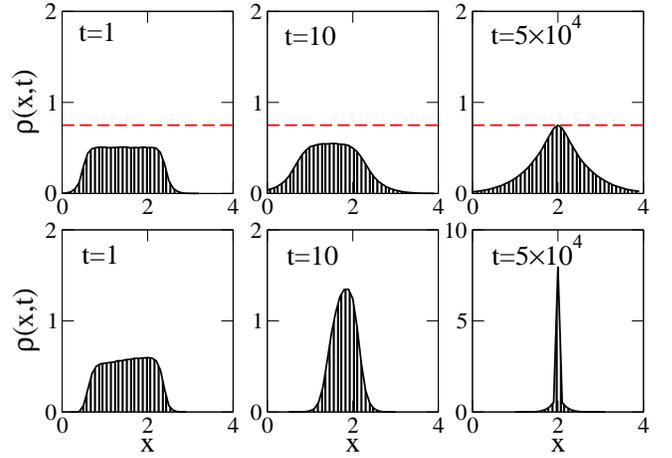}
}
\caption{Same as in Fig.\ $2$ but for quadratic external field $S(x)=\protect%
- \sigma (x-2)^2/2$. The upper row shows the formation of stationary
distribution for $\protect\sigma=0.64 < \protect\sigma_{cr}$. The bottom row
illustrates the density collapse that takes place for $\protect\sigma=10>
\protect\sigma_{cr} $. Here we consider parameters $A=4$ and $\protect\mu_0=4
$. Other parameters, initial and boundary conditions are the same as in
Fig.\ \ref{FIG2}. The critical value of $\protect\sigma$ is estimated numerically
to be $\protect\sigma_{cr} \simeq 0.65$. Dashed lines represent critical
value of the density $\protect\rho_{cr}=3/4$ given by Eq.\ (\protect\ref%
{rho_cr}). Notice that in this case the collapse takes place at $x=2$.}
\label{FIG3}
\end{figure}

We should note that the \textit{self-organized anomaly }is an universal
effect that can occur for any nonuniform external field $S(x).$ To
demonstrate that the boundary effects are irrelevant and to show that the
SOA does not depend on the form of $S(x)$, in particular on the derivative
of $S(x)$ at the point $x_m$, we consider quadratic external field with the
minimum at the center of the domain $[0,4]$
\begin{equation}
S(x)=- \sigma (x-2)^2/2.
\label{quadratic}
\end{equation}%
Here $\sigma>0$ is the strength parameter. Fig.\ \ref{FIG3} illustrates the phenomenon
of the density collapse that takes place at the point $x_{m}=2.$ This shows
that SOA is not a boundary effect.

\section{ Nonlinear Markovian model}

We should stress the fact that the self-organized anomalous regime occurs
only for the non-Markovian case. In this section we show that there is no
anomalous collapse for the Markovian nonlinear dynamics when the escape rate
$\mathbb{T}$ is independent of the residence time $\tau $%
\begin{equation}
\mathbb{T}(\rho )=\frac{\mu _{0}}{\tau _{0}\left( 1+A\rho \right) }.
\label{ind}
\end{equation}
We start with the Markovian master equation for the total density $\rho$. Using 
Eq.\ (\ref{masterGEN}), we have
\begin{eqnarray}
\frac{\partial \rho }{\partial t} &=&-\mathbb{T}(\rho )\rho (x,t)+p_{+}(x-a)%
\mathbb{T}\left( \rho(x-a,t) \right) \rho (x-a,t)  \notag \\
&&+p_{-}(x+a) \mathbb{T}\left( \rho (x+a,t) \right) \rho (x+a,t),
\label{master}
\end{eqnarray}%
where $p_{\pm }(x)$ is defined by (\ref{ppppp}). We use the Taylor series in
(\ref{master}) expanding the right hand side in the small $a$ and truncate
the series at the second term. The equation for $\rho $ takes the form
\begin{equation}
\frac{\partial \rho }{\partial t}=2\beta \frac{\partial }{\partial x}\left[
\frac{\partial S}{\partial x}D\left( \rho \right) \rho \right] +\frac{%
\partial ^{2}}{\partial x^{2}}\left[ D\left( \rho \right) \rho \right]
\label{non}
\end{equation}%
with the nonlinear diffusion coefficient%
\begin{equation}
D\left( \rho \right) =\frac{a^{2}\mathbb{T}(\rho )}{2}=\frac{a^{2}\mu _{0}}{%
2\tau _{0}\left( 1+A\rho \right) }.  \label{DD}
\end{equation}%
There is no anomalous collapse in this model for any form of the signalling
concentration $S(x)$. The non-uniform stationary solution of (\ref{non})
with (\ref{DD}) gives us a population aggregation profile. In the linear
case $A=0$ we have a classical Fokker--Planck equation involving the
external field $S(x).$ We should note that in this paper we do not consider
the aggregation process due to negative diffusion coefficient \cite{Turchin}%
. This effect might occur for some negative dependence of $\ $the escape
rate $\mathbb{T}(\rho )$ on the density $\rho.$ For our model with (\ref%
{ind}) the last term in Eq. (\ref{non}) can be modified in a such way that (%
\ref{non}) takes the form
\begin{equation*}
\frac{\partial \rho }{\partial t}=2\beta \frac{\partial }{\partial x}\left[
\frac{\partial S}{\partial x}D\left( \rho \right) \rho \right] +\frac{%
\partial }{\partial x}\left[ D_{m}\left( \rho \right) \frac{\partial \rho }{%
\partial x}\right]
\end{equation*}%
with%
\begin{equation}
D_{m}\left( \rho \right) =\frac{a^{2}\mu _{0}}{2\tau _{0}\left( 1+A\rho
\right) ^{2}}>0.  \label{ddd}
\end{equation}%
Clearly, the modified nonlinear diffusion coefficient $D_{m}\left( \rho
\right) $ is positive.

\section{ The underling mechanism for the self-organized anomalous regime}

\subsection{Stationary aggregation profile}

Can we understand the underling mechanism for the \textit{self-organized
anomaly (SOA) }observed in our Monte Carlo simulations? They show that in
the self-organized anomalous regime, the standard stationary density profile
(aggregation pattern) does not exist. So, it is natural \ first to find a
stationary solution $\xi _{st}(x,\tau )$ to the Eq.\ (\ref{dif1}) from
\begin{equation*}
\frac{\partial \xi _{st}}{\partial \tau }=-\mathbb{T}(\tau ,\rho _{st}\left(
x\right) )\xi _{st}.
\end{equation*}%
Using the escape rate (\ref{d}) we obtain
\begin{equation*}
\xi _{st}(x,\tau )=\xi _{st}(x,0)\left( \frac{\tau _{0}}{\tau _{0}+\tau }%
\right) ^{\frac{\mu _{0}}{1+A\rho _{st}(x)}}.
\end{equation*}%
This can be rewritten in terms of the \textit{density-dependent }power law
survival function
\begin{equation}
\Psi (\tau ,\rho _{st})=\left( \frac{\tau _{0}}{\tau _{0}+\tau }\right) ^{%
\frac{\mu _{0}}{1+A\rho _{st}(x)}}  \label{ss}
\end{equation}%
and the stationary arrival rate $j_{st}(x)=\xi _{st}(x,0)$ as
\begin{equation}
\xi _{st}(x,\tau )=j_{st}(x)\Psi (\tau ,\rho _{st}).  \label{den}
\end{equation}%
In the stationary case the arrival rate of particles $j_{st}(x)$ to the
point $x$ and the escape rate of particles from the point $x$ are the same: $%
j_{st}(x)=i_{st}(x)$. The stationary density $\rho _{st}(x)$ can be obtained
from (\ref{den5}), (\ref{den}) in the limit $t\rightarrow \infty$:%
\begin{equation}
\rho _{st}(x)=\int_{0}^{\infty }\xi _{st}(x,\tau )d\tau
=i_{st}(x)\int_{0}^{\infty }\Psi \left( \tau ,\rho _{st}\right) d\tau .
\label{ku}
\end{equation}%
Note that
\begin{equation}
\bar{T}\left( \rho _{st}(x)\right) =\int_{0}^{\infty }\Psi (\tau ,\rho
_{st})d\tau  \label{ttt}
\end{equation}%
can be interpreted as the expected value of the random residence time $T$
whose survival function is given by (\ref{ss}). It should be emphasized that
we cannot introduce the power-law survival function as the function of
non-stationary population density $\rho (x,t).$ It follows from (\ref{ku})
and (\ref{ttt}) that the stationary escape rate $i_{st}(x)$ can be written
in the standard Markovian form
\begin{equation}
i_{st}(x)=\frac{1}{\bar{T}\left( \rho _{st}(x)\right) }\rho _{st}(x).
\label{ii}
\end{equation}%
Using Eq.\ (\ref{masterGEN}), we write the stationary master equation:
\begin{equation}
-i_{st}(x)+p_{+}(x-a)i_{st}(x-a)+p_{-}(x+a)i_{st}(x+a)=0.  \label{Master}
\end{equation}%
Using probabilities (\ref{ppppp}), in the limit $a\rightarrow 0$, we obtain
from (\ref{ii}) and (\ref{Master}) the nonlinear stationary
advection-diffusion equation for the population density $\rho _{st}(x):$
\begin{equation}
2\beta \frac{\partial }{\partial x}\left[ \frac{S^{\prime }(x)\rho _{st}(x)}{%
\bar{T}\left( \rho _{st}(x)\right) }\right] +\frac{\partial ^{2}}{\partial
x^{2}}\left[ \frac{\rho _{st}(x)}{\bar{T}\left( \rho _{st}(x)\right) }\right]
=0.  \label{ssss}
\end{equation}%
It follows from (\ref{ssss}) that the steady profile $\rho _{st}(x)$ on the
interval $\left[ 0,L\right] $ with the reflecting boundaries can be found
from the non-linear equation $\rho _{st}(x)=N^{-1}\bar{T}\left( \rho
_{st}\right) \exp \left[ 2\beta S(x)\right] $, where $N$ is determined by
the normalization condition $N=\int_{0}^{L}\bar{T}\left( \rho _{st}\left(
x\right) \right) \exp \left[ 2\beta S(x)\right] dx.$ This stationary profile
$\rho _{st}(x)$ is illustrated in Fig.\ \ref{FIG2} for the linear external field Eq.\ (\ref{linear}) and 
in Fig.\ \ref{FIG3} for quadratic field Eq.\ (\ref{quadratic}) (see right most profiles in
the top rows).

\begin{figure}[t]
\vspace{10pt} \centerline{
\psfig{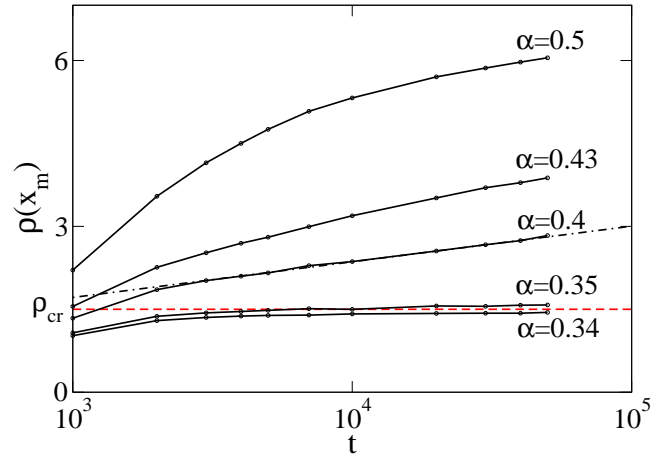}
} \vspace*{30pt}
\caption{Dependence of the population density at the boundary $x_{m}=4$, $%
\protect\rho (x_{m},t)$, for linear field $S(x)=\protect\alpha x$.
Parameters are the same as in Fig.\ \protect\ref{FIG2}. Dashed line mark
critical values of $\protect\rho_{cr} =1.5$ given by Eq.\ (\protect\ref%
{rho_cr}). Dashed-dotted curve is the best fit for intermediate value of $%
\protect\alpha =0.4$ by logarithmic function $\protect\rho (x_{m},t)\sim
0.28\ln (t)-0.22$ . }
\label{FIG4}
\end{figure}

\subsection{Conditions for self-organized anomaly} 

The question arises why the increase of the parameter $\alpha $ for the linear external field (\ref{linear}) or $\sigma$ 
for the quadratic field Eq.\ (\ref{quadratic}) above the corresponding critical value $\alpha _{cr}$ or $\sigma_{cr}$ 
leads to a collapse of stationary
aggregation pattern (see bottom rows in Figs. \ref{FIG2} and \ref{FIG3}). Our main idea is
that when $\alpha $ or $\sigma$ gets bigger it leads to the increase of the population
density at the point $x_{m}$ where density $\rho _{st}(x)$ takes the maximum
value and the divergence of the integral
\begin{equation}
\int_{0}^{\infty }\Psi (\tau ,\rho _{st}(x_{m}))d\tau .  \label{con}
\end{equation}%
In another words: the \textit{self-organized anomaly} occurs when
the effective mean residence time $\bar{T}$ $(x_{m})=\int_{0}^{\infty }\Psi
(\tau ,\rho _{st}(x_{m}))d\tau $ becomes infinite and the stationary Eq. (%
\ref{ssss}) breaks down. The reason why we call this regime anomalous is
that the divergence of the mean waiting time explains anomalous subdiffusive
behavior of the random walkers \cite{Klafter,Kla,Men2}. The essential
difference to the standard CTRW theory is that we use the stationary
density-dependent power law survival function. Although SOA is similar to
the phenomenon of anomalous aggregation \cite{Fed3} or the accumulation of
subdiffusive particles in one of two infinite domains with two different
values of anomalous exponents \cite{Kor}, it is essentially different.
Anomalous conditions are not imposed by power law waiting time PDF (\ref%
{wtden}) with the anomalous exponent $\mu _{0}<1$.\ \textit{Anomalous regime
is self-organized} for $\mu _{0}>1$ through the nonlinear interactions of
random walkers due to social crowding effects described by (\ref{T}).

Substitution of the survival function (\ref{ss}) into (\ref{con}) gives
\begin{equation}
\bar{T}(x_{m})=\frac{\tau _{0}\left[ 1+A\rho _{st}(x_{m})\right] }{\mu
_{0}-1-A\rho _{st}(x_{m})}.
\end{equation}%
The divergence of $\bar{T}(x_{m})$ gives the critical density $\rho_{cr}$:
\begin{equation}
\rho_{cr}=\frac{\mu _{0}-1}{A}.  \label{rho_cr}
\end{equation}
One can also write the critical condition as $\gamma=1$, where
\begin{equation}
\gamma =\frac{\mu _{0}}{1+A\rho_{st}(x_{m})},\quad \mu _{0}>1.
\end{equation}%
In particular, for the linear external field $S(x)=\alpha x$ in the interval
$\left[ 0,4\right] $ the stationary density $\rho _{st}(x)$ has a maximum
value at the point $x_{m}=4$. We can find the critical value $\alpha _{cr}$
as
\begin{equation}
\lim_{\alpha \rightarrow \alpha _{cr}}\frac{\mu _{0}}{1+A\rho _{st}(x_{m})}%
=1,\quad \mu _{0}>1.
\end{equation}%

Numerical simulations\ presented in Fig.\ \ref{FIG2} and \ref{FIG3} are in excellent
agreement with Eq. (\ref{rho_cr}). In Fig.\ \ref{FIG4} we illustrate the time
evolution of the population density $\rho (x_{m},t)$ at the boundary $x_{m}=4
$ for linear field $S(x)$ with different strength of the signalling $\alpha.$ Transition to
self-organized anomalous regime is observed by the transition of the density
through the critical value given by Eq. (\ref{rho_cr}). For intermediate
values of $\alpha $, we find that the density grows as $\ln t$ see Fig.\ \ref{FIG4}.

\subsection{The role of the initial conditions.}

\begin{figure}[t]
\centerline{
\psfig{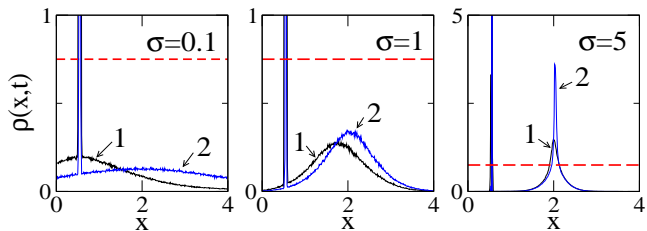}
}
\caption{The role of the initial conditions for the quadratic external field $S(x)$ with different field strength $\sigma$. We have considered parameters $\tau_0=1$, $\beta=1$, $A=4$, $\mu_0=4$, $L=4$ and $M=400$, so that $a=0.01$. In contrast to Fig.\ \ref{FIG3}, the initial distribution of particles is chosen to violate the condition in Eq.\ (\ref{condition}). We have used ensemble of $%
10^{6}$ walkers uniformly distributed at time $t=0$ on the interval $%
[0.5,0.6]$ with zero residence time ($\protect\tau =0$). The boundaries at $x=0$ and $x=4$ are assumed to be reflecting. The dashed lines correspond to the critical value of the density $\rho_{cr}=3/4$, see Eq.\ (\ref{rho_cr}). Curve $1$ corresponds to $t=10^4$ and curve $2$ to $t=7 \times 10^4$.}
\label{FIG5}
\end{figure}

In our numerical simulations we have used the initial conditions for which
all walkers have zero residence time (no aging effects) and the maximum
value of the initial density $\rho _{0}(x)$ obeys the inequality
\begin{equation}
\max \rho_0(x) < \rho_{cr},
\label{condition}
\end{equation}%
where $\rho_{cr}$ is given by Eq.\ (\ref{rho_cr}). However, when this condition is violated the SOA regime could have different scenarios. In Fig.\ \ref{FIG5} some preliminary results of numerical simulations are shown. We observe three scenarios depending on the strength of the external field $\sigma$. The left most panel shows the anomalous aggregation of the walkers in the region of non-zero initial density $[0.5, 0.6]$. The peak of the density in this region is infinitely increasing (this increase is not captured in Fig.\ \ref{FIG5} , only the peak is visible). The density in the other region reaches a quasi stationary form approximately represented by curve $2$ for $t=7 \times 10^4$. Note, however, that it is not a true stationary state since the particles continue to accumulate in the interval $[0.5, 0.6]$. The middle panel demonstrates similar scenario for $\sigma=1$. However, now the field is strong enough to temporarily accumulate particles at the critical point$x_m=2$. Our assumption is that similarly to the previous case, in the long time limit these particle will be concentrated in the region of non-zero initial conditions. A different scenario occur for large strength of the external field. The right most panel shows this situation for $\sigma=5$. In this case the external field is able to strongly attract particles to its critical point $x_m=2$. At some time the density at this point exceed the critical value Eq.\ (\ref{rho_cr}). Therefore, in this case the anomalous aggregation of the density is observed at two places, at the region of non-zero initial conditions and at the critical point $x_m=2$. We call this phenomenon as transient anomalous bi-modal aggregation.

\section{Conclusions}

We have discovered the phenomenon of\textit{\ self-organized anomaly} (SOA)
which takes place in a population of particles performing nonlinear and
non-Markovian random walk. This random walk involves social crowding effects
for which the dispersal rate of particles is a decreasing function of the
population density and residence time \cite{Erik,Men3}. Monte Carlo
simulations shows that the regime of \textit{self-organized anomaly } leads
to a collapse of a stationary aggregation pattern when all particles
concentrate inside a tiny region of space and form a non-stationary high
density cluster. The maximum population density slowly increases with time
as $\ln t.$ We should note that the anomalous regime is self-organized and
arises spontaneously without the need for introduction of the power law
waiting time distribution with infinite mean time. Only in a stationary case
one can obtain a power-law density-dependent survival function and define
the critical condition as the divergence of mean residence time. SOA gives a
new possible mechanism for chemotactic collapse in a population of living
organisms as an alternative to the celebrated Patlak-Keller-Segel theory
\cite{sad2,Brenner,The}. The crossover from the standard stationary
aggregation pattern to a non-stationary anomalous aggregation as the
strength of chemotactic force increases can be interpreted as
non-equilibrium phase transition. Our theory can be used to explain various
anomalous aggregation phenomenon including accumulation of phagotrophic
protists in attractive patches where they become almost immobile \cite{Pro}.
It would be interesting (i) to analyze extensions of our model by taking
into account additional effects like volume feeling preventing unlimited
density growth \cite{Fed2} and (ii) apply our model for the analysis of how
self-organization and an anomalous cooperative effect arise in social
systems \cite{S}.

This work was funded by EPSRC grant EP/J019526/1. The authors thank Steve
Falconer for the helpful suggestions.

\end{document}